\documentclass[journal=jpcc,manuscript=article]{achemso}

\usepackage[version=3]{mhchem} % Formula subscripts using \ce{}

\usepackage{achemso}

\author{Jing Wang$^{a,b}$}
\author{Hong-Man Ma,$^{a}$}
\author{Ying Liu$^{a,c}$}
\email{yliu@hebtu.edu.cn}

\affiliation
{$^{a}$Department of Physics and Hebei Advanced Thin Film Laboratory, Hebei Normal University,\\ Shijiazhuang 050024, Hebei, China.\\
$^{b}$State Key Laboratory for Superlattices and Microstructures, Institute of Semiconductors, \\Chinese Academy of Sciences, Beijing 100083, China.\\
$^{c}$National Key Laboratory for Materials Simulation and Design, Beijing 100083, China}

\title{Two-Dimensional Scandium Carbide Monolayer and its Nanotubes}

\begin{document}

\begin{abstract}
A two-dimensional scandium carbide monolayer with a Sc$_3$C$_{10}$ primitive cell (Sc$_3$C$_{10}$ sheet) has been identified using first-principles density functional theory. In the Sc$_3$C$_{10}$ sheet, there is a similar basic structure to the one in the \textit{Volleyballene} Sc$_{20}$C$_{60}$, the Sc$_8$C$_{10}$ subunit, in which two connected carbon pentagons are surrounded by one scandium octagon. The hybridization between Sc $d$ orbitals and C $s$-$p$ orbitals is crucial for stabilizing the Sc$_3$C$_{10}$ sheet. \textit{Ab initio} molecular dynamics simulations demonstrate that this Sc$_3$C$_{10}$ sheet is exceptionally stable. In addition, a series of stable ScC nanotubes have been obtained by rolling up this Sc$_3$C$_{10}$ sheet. All nanotubes studied have been found to be metallic.
\end{abstract}

Graphene and transition metal dichalcogenides (TMDCs), the two major types of two-dimensional (2D) materials have been the object of intense investigations as candidate materials for future nanoelectronics applications \cite{1,+1,2,+2,3,4}. Developments in the field such as a field-effect transistor (FET) require a moderate electronic band gap, a reasonably high carrier mobility, and excellent electrode-channel contacts\cite{+1,+3}. Graphene, the most famous member of the 2D material family, possesses remarkable electronic and mechanical properties, but the lack of a native band gap severely limits its applications in nanotransistors.\cite{5} Recent great effort has been directed toward opening a gap in various graphene-based nanostructures. Nonetheless the devices designed all have a low "on-off" current ratio\cite{3}. Unlike graphene, monolayer molybdenum disulfide (MoS$_2$), a representative of the 2D TMDCs, does have a direct band gap of $\sim$1.8 eV\cite{10}, but a carrier mobility of only about 200 cm$^2$V$^{-1}$s$^{-1}$\cite{4}, which is not sufficiently high for many applications\cite{6} Thus, the zero bandgap of graphene and the relatively low mobility in TMDCs limit their applications for nanoelectronics.

Very recently, monolayers of the element phosphorus (phosphorene), have been mechanically exfoliated from bulk black phosphorus and found to be a natural $p$-type semiconductor with an appreciable, direct band gap\cite{7,8,9}. This exciting result has encouraged continued searches for novel 2D materials. In the present work, we propose a novel 2D monolayer, based on the basic unit of the \textit{Volleyballene} Sc$_{20}$C$_{60}$\cite{+4}, with robust stability and excellent structural and physical properties: a monolayer, scandium carbide sheet, referred to below as Sc$_3$C$_{10}$ sheet. First principles calculations also show that stable ScC nanotubes can be obtained from these sheets.

Our calculations were performed within the framework of spin-polarized density functional theory (DFT) with the generalized gradient approximation (GGA) using the exchange-correlation potential described by Perdue-Burke-Ernzerhof (PBE)\cite{11}. The calculations were carried out with unrestricted symmetry using a double-numerical polarized (DNP) basis set\cite{12}. For the transition metal atoms, relativistic effects in the core were included by using the DFT semi-core pseudopotentials (DSPP)\cite{13}. All structures were fully relaxed, and geometric optimizations were performed with convergence thresholds of 10$^{-5}$ hartree (Ha) for the energy, 2$\times$10$^{-3}$ Ha/{\AA} for forces, and 5$\times$10$^{-3}$ {\AA} for the atomic displacements. For the ScC nanotube calculations, the vacuum spaces between tubes were made larger than 10 {\AA} to avoid interactions between the tubes.

Figure 1$a$ shows the atomic structure of the most stable Sc$_3$C$_{10}$ sheet obtained in the structural search. The primitive cell contains 3 scandium atoms and 10 carbons with the chemical formula of Sc$_3$C$_{10}$. The lattice parameters are \textbf{\textit{a}}$_\textbf{1}$ = \textbf{\textit{a}}$_\textbf{2}$ = 8.855 {\AA} and $\alpha$ = 142$^{\circ}$, respectively. A unit cell (\textbf{\textit{b}}$_\textbf{1}$, \textbf{\textit{b}}$_\textbf{2}$), twice the size of the primitive cell, is also given in Fig. 1$a$. In the Sc$_3$C$_{10}$ sheet, there is a basic structure, the Sc$_8$C$_{10}$ subunit, highlighted in the top left corner of Fig. 1$a$. In the Sc$_8$C$_{10}$ subunit, there are two pentagonal rings made of carbon atoms (C-pentagons) and one octagonal ring of scandium atoms (Sc-octagons). It may be seen that every group of two C-pentagons is surrounded by one Sc-octagon. This new scandium carbide sheet may thus be viewed as consisting of Sc$_8$C$_{10}$ subunits set in a crisscross pattern. The average Sc-Sc bond length is 3.340 {\AA} with two distinct Sc-Sc bond lengths: 3.351 {\AA} along the horizontal direction (\textbf{\textit{b}}$_\textbf{1}$ in Fig. 1) and 3.328 {\AA} for the other cases. For the C-pentagons, there are three C-C double bonds (1.428 {\AA}) and two C-C single bonds (1.466 {\AA}). Along with a 1.437 {\AA} C-C bond connecting the two C-pentagons, the average C-C bond length is thus 1.443 {\AA}. The average Sc-C bond length is 2.299 {\AA}.

The stability of the Sc$_3$C$_{10}$ sheet was investigated by analyzing its bond characteristics, and confirmed using \textit{ab initio} molecular dynamics simulations. Figure 1$b$ shows the deformation electron density, which reveals electron transfer from the Sc atoms to the C atoms. Mulliken population analysis shows a charge transfer of $\sim$0.6$e$ from each Sc atom, mainly from the Sc 3$d$ state. For the C atoms, there are obvious characteristics of $sp^2$-like hybridization, and each C atom has three $\sigma$ bonds. For the Sc atoms there are obvious $d$ orbital characteristics with four lobes. The Sc atom in the middle of the primitive cell bonds with the four neighboring C atoms through its $d$ orbital. For the remaining two Sc atoms of the primitive cell, each Sc interacts with the two C atoms which are more centrally located than are the other six carbons. Close examination of the partial density of states (PDOS) as shown in Figure 1$c$ further confirms the hybridization between the Sc $d$ orbitals and C $s$-$p$ orbitals. This is crucial for stabilizing the planar Sc$_3$C$_{10}$ sheet.

Next, \textit{ab initio} molecular dynamics simulations with an NVE ensemble were carried out with a time step of 1.0 $fs$. The results of the NVE dynamic simulations indicated that the Sc$_3$C$_{10}$ sheet retained its original topological structure and was not disrupted over the course of a 2.0 $ps$ dynamic simulation at an initial temperature of 1400 K. The Sc$_3$C$_{10}$ sheet therefore has good thermodynamic stability. In addition, analysis of the band structure (\textit{see} Fig. 1$c$) shows a direct band gap $\sim$0.62 eV at the GGA/PBE level.

Just as graphene is the precursor of carbon nanotubes, a series of ScC nanotubes with different diameters and chiralities could be constructed based on the Sc$_3$C$_{10}$ sheet. We first illustrate how to describe these ScC nanotubes. Because the Sc$_3$C$_{10}$ sheet has low symmetry, it is really not appropriate to classify the ScC nanotubes using the primary vectors (\textbf{\textit{a}}$_\textbf{1}$, \textbf{\textit{a}}$_\textbf{2}$) of the orthorhombic lattice. In the present case, the lattice vectors of the rectangular lattice,  \textbf{\textit{b}}$_\textbf{1}$ and \textbf{\textit{b}}$_\textbf{2}$ (as shown in Fig. 1$a$), seem to be more appropriate and convenient for labeling ScC nanotubes with integer multiples of the basis vectors.

After geometry optimization, it was found that the (\textit{p}, 0) tubes had all collapsed. Only the (0, \textit{q}) nanotubes were stable. Calculation were performed on these (0, \textit{q}) nanotubes with \textit{q} = 2, 3, 4, 5 and diameters in the range 1.83-4.53 \AA, and the stabilities and electronic properties were explored. Figure 2 shows the binding energy per atom of the (0, \textit{q}) nanotubes \textit{vs} the diameter. It can be seen that the binding energy approaches that of the corresponding Sc$_3$C$_{10}$ sheet as the diameter increases. The (0, \textit{q}) ScC nanotubes with large diameter have relatively high stability and the stability increases with the diameter. Analysis of the electronic structures of the (0, \textit{q}) ScC nanotubes indicates that all four (0, \textit{q}) tubes rolled from the Sc$_3$C$_{10}$ sheet are metallic. The band structures and densities of states (DOS) of the (0, \textit{q}) ScC nanotubes are shown in Figure 3. Close examination of the band structures indicates that the (0, 2) nanotube is different from the other three examined. For the latter cases, all of the (0, 3), (0, 4), and (0, 5) nanotubes exhibit a gap slightly above the fermi level. All three band gaps are direct band gaps at the $\Gamma$-point and the gap sizes increase as the diameter increases. The band gaps are $\sim$0.60, 0.64, and 0.71 eV for the (0, 3), (0, 4), and (0, 5) nanotubes, respectively. The band structure of the (0, 2) tube, on the other hand, shows several bands in the vicinity of the fermi level, which ensures a large carrier density.

In conclusion, our first-principles studies have identified a stable Sc$_3$C$_{10}$ sheet using both static and dynamic \textit{ab initio} calculations. The new scandium carbide sheet may be viewed as consisting of Sc$_8$C$_{10}$ subunits arranged in a crisscross pattern. Hybridization between Sc $d$ orbitals and C $s$-$p$ orbitals is essential for stabilizing the Sc$_3$C$_{10}$ sheet. In addition, all the stable ScC nanotubes rolled from this Sc$_3$C$_{10}$ sheet were found to be metallic within the scope of the approximations used in our research. This prediction is expected to motivate experimental efforts in view of the fundamental value and potential applications of these structures.

\acknowledgement
The authors thank Dr. N. E. Davison for his help with the language. This work is supported by the National Natural Science Foundation of China (Grant Nos. 11274089, 11304076, and U1331116), the Natural Science Foundation of Hebei Province (Grant Nos. A2012205066 and A2012205069), and the Science Foundation of Hebei Education Award for Distinguished Young Scholars (Grant No. YQ2013008). We also acknowledge partially financial support from the 973 Project in China under Grant No. 2011CB606401.

\clearpage

\begin{figure*}
 \includegraphics[scale=1.15]{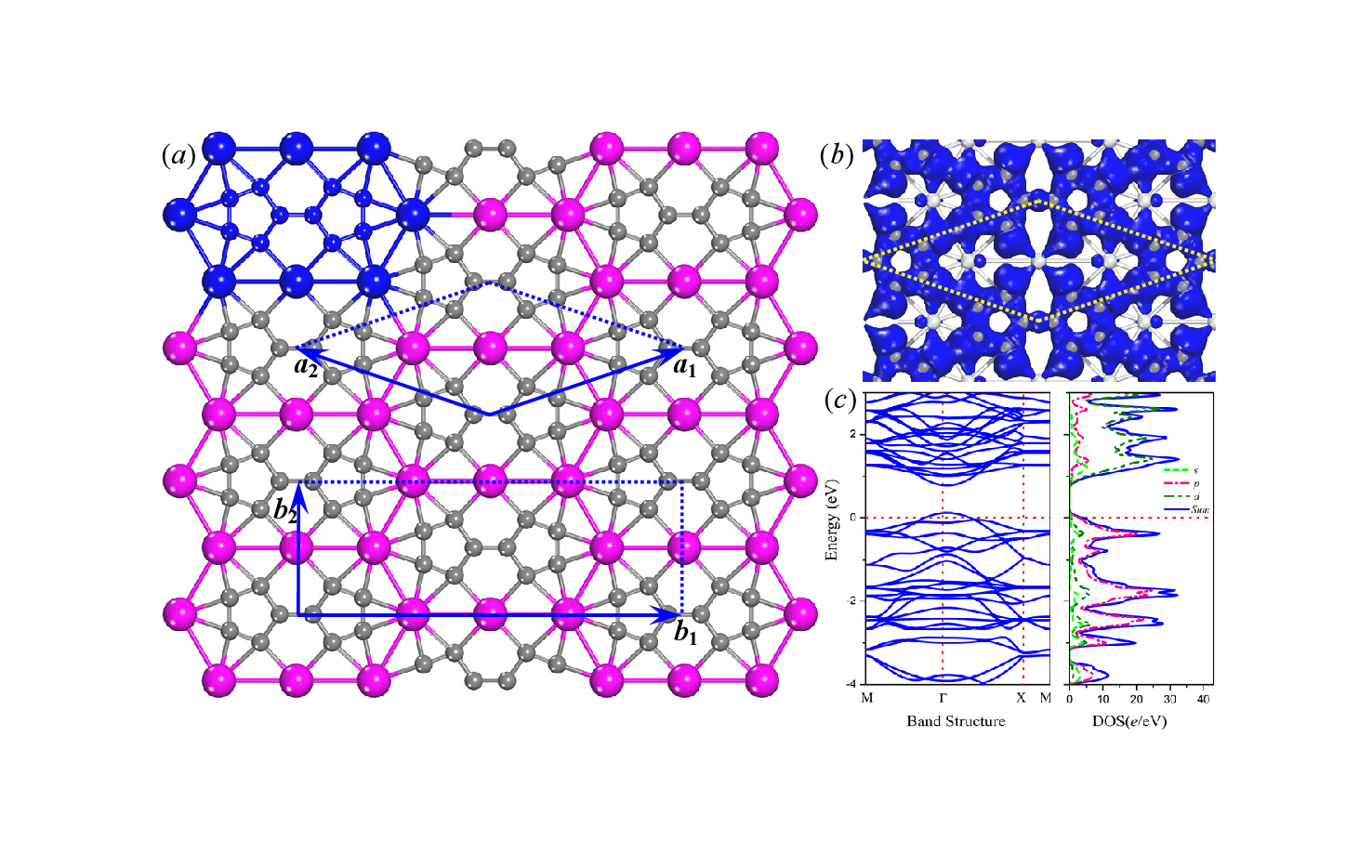}
  \caption{(\textit{a}) The configuration, (\textit{b}) the deformation electron density, and (\textit{c}) the band structure and partial density of states (PDOS) of the Sc$_3$C$_{10}$ sheet. The large and small balls represent Sc and C atoms, respectively. The blue balls at the top left corner show the basic Sc$_8$C$_{10}$ subunit. The directed lines (\textit{\textbf{a}}$_\textbf{1}$, \textit{\textbf{a}}$_\textbf{2}$ and \textit{\textbf{b}}$_\textbf{1}$, \textit{\textbf{b}}$_\textbf{2}$) represent the lattice vectors as described in the text. The iso-value of ($b$) is set to 0.01 e/\AA$^3$.}
\end{figure*}

\begin{figure*}
\includegraphics[scale=0.55]{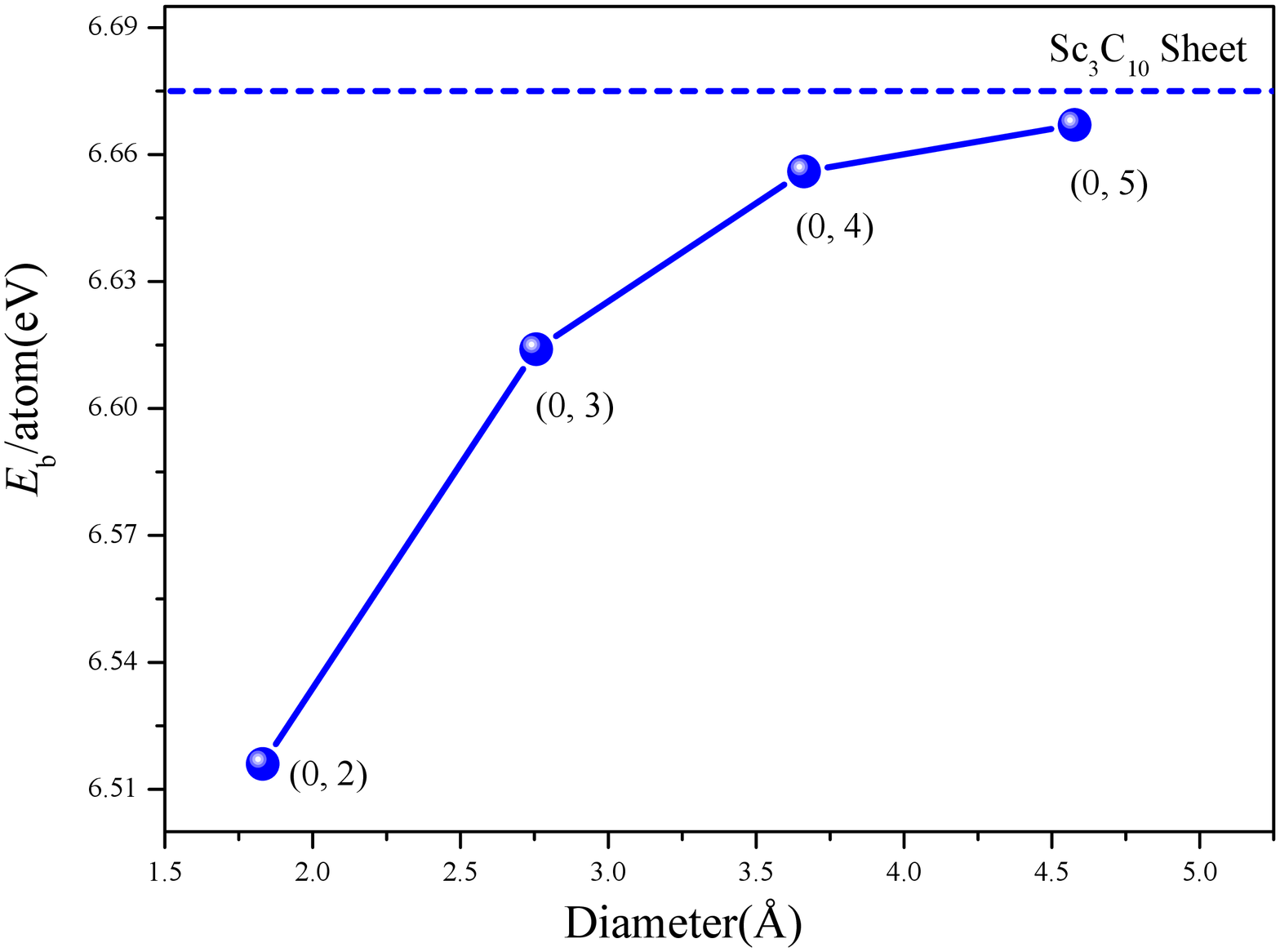}
  \caption{The binding energy per atom of ScC nanotubes \textit{vs} the diameter. The labels describe the nanotubes in terms of the rectangular lattice vectors of the Sc$_3$C$_{10}$ sheet as outlined in the text.}
\end{figure*}

\begin{figure*}
 \includegraphics[scale=0.55]{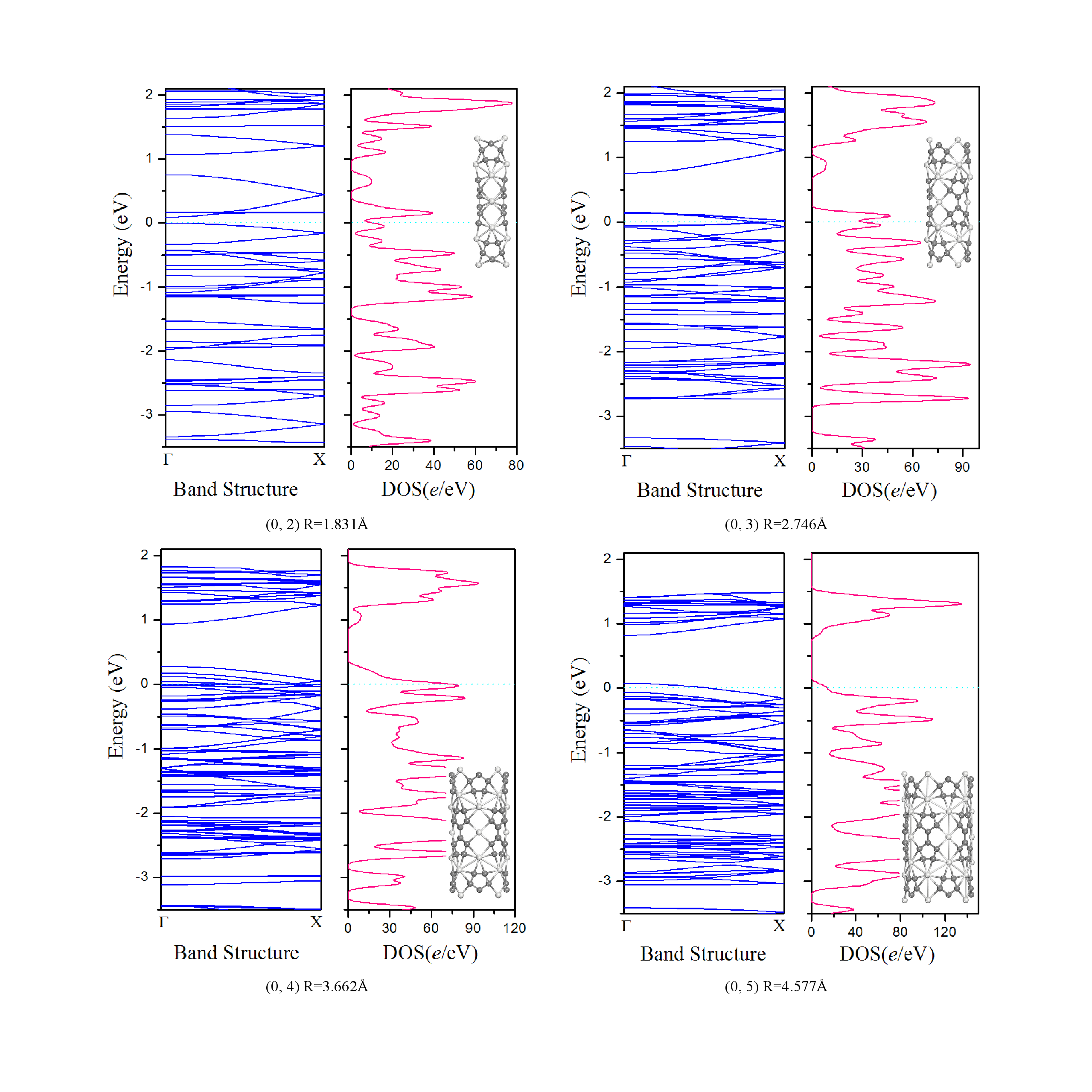}
  \caption{The band structures and densities of states (DOS) for (0, \textit{q}) nanotubes with \textit{q} = 2, 3, 4, 5, as well as the schematics of corresponding ScC nanotubes.}
\end{figure*}

\end{document}